\documentclass[preprint,aps,floats,nofootinbib,amssymb,showpacs,amsmath]{revtex4}
\usepackage{epsf,epsfig}

\newcommand{\be}{\begin{equation}}
\newcommand{\ee}{\end{equation}}
\newcommand{\bea}{\begin{eqnarray}}
\newcommand{\eea}{\end{eqnarray}}
\newcommand{\nn}{\nonumber}
\newcommand{\ba}{\begin{array}} 
\newcommand{\ea}{\end{array}}

\newcommand{\lsim}{
\mathrel{\hbox{\rlap{\hbox{\lower4pt\hbox{$\sim$}}}\hbox{$<$}}}}
\newcommand{\gsim}{
\mathrel{\hbox{\rlap{\hbox{\lower4pt\hbox{$\sim$}}}\hbox{$>$}}}}

\preprint{
\hbox to \hsize{
\hfill$\vcenter{
        \hbox{\bf UFIFT-HEP-07-09}
            \hbox{\bf FSU-HEP-070705}
                \hbox{September 2007}}$}
}

\begin{document}

\title{
\vspace*{.75in}
$\boldsymbol{U(1)'}$ solution to the $\boldsymbol{\mu}$-problem and 
the proton decay problem in supersymmetry without $R$-parity
}

\author{
Hye-Sung Lee$^1$, Konstantin T. Matchev$^1$, and Ting T. Wang$^2$}

\affiliation{
$^1$Institute for Fundamental Theory, University of Florida,
Gainesville, FL 32611 \\
$^2$Department of Physics, Florida State University,
Tallahassee, FL 32306
\vspace*{.5in}}

\thispagestyle{empty}
\begin{abstract}
\noindent
The Minimal Supersymmetric Standard Model (MSSM) is plagued by two 
major fine-tuning problems: the $\mu$-problem and the proton decay problem.
We present a simultaneous solution to both problems
within the framework of a $U(1)'$-extended  
MSSM (UMSSM), without requiring $R$-parity conservation.
We identify several classes of phenomenologically viable models 
and provide specific examples of $U(1)'$ charge assignments.
Our models generically contain {\em either} lepton number violating
{\em or} baryon number violating renormalizable interactions, 
whose coexistence is nevertheless automatically forbidden by the 
new $U(1)'$ gauge symmetry. 
The $U(1)'$ symmetry also prohibits the potentially 
dangerous and often ignored higher-dimensional proton decay operators 
such as $QQQL$ and $U^cU^cD^cE^c$ which are still allowed by $R$-parity. 
Thus, under minimal assumptions, we show that
once the $\mu$-problem is solved, the proton is sufficiently stable,
even in the presence of a minimum set of exotics fields, 
as required for anomaly cancellation.
Our models provide impetus for pursuing 
the collider phenomenology of $R$-parity violation within 
the UMSSM framework.
\end{abstract} 

\maketitle

\newpage

\section{Introduction}

Supersymmetry (SUSY) at the Terascale has been the leading candidate for physics 
beyond the Standard Model (SM). We do not know the concrete manifestation of 
supersymmetry at low energies, but the Minimal Supersymmetric Standard Model (MSSM)
already incorporates most of the advantages of supersymmetry and 
has proved to be a useful playground for investigations of the possible SUSY
signatures at high energy colliders such as the Tevatron and the Large Hadron Collider (LHC). 
In spite of its successes, however, the MSSM does not exhaust all 
possibilities and, given its shortcomings discussed below, 
it is certainly worth pursuing alternative, more general low-energy 
supersymmetric theories.

One of the most celebrated successes of low-energy supersymmetry is the 
resolution of the gauge hierarchy problem of the SM. SUSY protects 
the Higgs mass and the associated electroweak scale from the dangerous 
quadratically divergent radiative corrections. However, the MSSM itself suffers 
from its own fine-tuning problems. First, there is the so-called $\mu$-problem \cite{muproblem}, 
which is associated with the following superpotential
coupling of the two MSSM Higgs doublets $H_1$ and $H_2$:
\be
W_{\mu} = \mu H_2 H_1\ .
\label{eq:mu}
\ee
Since this coupling is allowed by both supersymmetry and gauge symmetry,
there is no natural (i.e. in terms of a symmetry) explanation, at least 
within the MSSM, as to why the value of the $\mu$ parameter is so much 
smaller than the fundamental (Planck or string) scale.
To fix this problem in a natural way, one has to introduce a symmetry
which would prohibit the original $\mu$ term (\ref{eq:mu}). 
However, 
in the end this symmetry needs to be broken, since
a vanishing $\mu$ term would imply very light charginos, 
in violation of the LEP search limits \cite{LEPchargino}.
Therefore, a viable model should dynamically generate an effective $\mu$ term. 
This is typically done by introducing a Higgs singlet $S$ coupling to the 
MSSM Higgs doublets as
\be
W_{\mu_{\rm eff}} = h S H_2 H_1\ .
\label{eq:mueff}
\ee
The singlet $S$ is charged under the new symmetry, so that the
original $\mu$ term (\ref{eq:mu}) is forbidden. The vacuum expectation value 
(VEV) of $S$ would then break the symmetry and 
play the role of an effective $\mu$ parameter. 
Depending on the type of the new symmetry, 
the models can be classified into several categories \cite{CPNSH}.
For instance, when the symmetry is a $\boldsymbol{Z_3}$ discrete symmetry, 
one obtains the Next-to-MSSM (NMSSM) \cite{NMSSM}, when the symmetry is an Abelian 
gauge symmetry $U(1)'$, we have the $U(1)'$-extended MSSM (UMSSM) \cite{UMSSM}, etc.
(Other options include the Minimal Nonminimal 
SSM (MNSSM) \cite{nMSSM} and the Essential SSM (ESSM) \cite{ESSM}.)
In this study we shall work within the UMSSM framework, and we shall use
the additional $U(1)'$ gauge interaction to forbid the original $\mu$ term (\ref{eq:mu})
while allowing the effective $\mu$ term (\ref{eq:mueff}).
We shall completely specify the particle content of the model
and will demand that the new $U(1)'$ gauge symmetry is non-anomalous.
An extra $U(1)$ symmetry is supported by many new physics paradigms including 
grand unified theories \cite{U1_gut1, U1_gut2}, extra dimensions \cite{U1_extradim}, 
superstrings \cite{U1_string}, little Higgs \cite{U1_littleHiggs}, 
dynamical symmetry breaking \cite{U1_strongdynamics}, 
and Stueckelberg mechanism \cite{U1_stueckelberg}.

The other fine-tuning problem of the MSSM is related to the existence of 
lepton number violating (LV) terms
\be
W_{\rm LV} = \mu'_i H_2 L_i + \lambda_{ijk} L_i L_j E^c_k + \lambda'_{ijk} L_i Q_j D^c_k 
\label{eq:LV}
\ee
and baryon number violating (BV) terms 
\be 
W_{\rm BV} = \lambda''_{ijk} U^c_i D^c_j D^c_k
\label{eq:BV}
\ee
in the superpotential. Here $i,j,k$ are generation indices and summation 
over repeated indices is implied. The couplings (\ref{eq:LV}) and (\ref{eq:BV})
are again allowed by all gauge symmetries and supersymmetry and may even occur in the underlying grand unified theory \cite{Rizzo:1992ts}. The presence 
of both types of such terms would lead to unacceptably rapid proton decay 
unless certain combinations of couplings are tuned to be extremely small 
($\lambda \lambda''\lsim 10^{-21}$, $\lambda' \lambda''\lsim 10^{-27}$ \cite{RPVMSSM}). 
The standard practice for dealing with this fine-tuning problem is again to impose a new 
symmetry, the so-called $R$-parity \cite{Rparity}, which is the only 
other new symmetry in the MSSM besides supersymmetry. $R$-parity forbids 
{\em both} types of problematic terms (\ref{eq:LV}) and (\ref{eq:BV})
and the proton appears to be safe. 

At this point one might question whether it was really necessary to forbid
{\em both} (\ref{eq:LV}) and (\ref{eq:BV}). Indeed, since proton decay 
requires both LV and BV interactions, forbidding either of them 
would be sufficient to stabilize the proton. In this sense, the imposition of 
$R$-parity is far from being the minimalist approach, since it eliminates a 
large chunk of potentially interesting phenomenology related to the physics of 
$R$-parity violation (RPV) \cite{Barger}. In this study, we shall therefore
utilize the $U(1)'$ gauge symmetry to forbid some, but not all $R$-parity 
violating interactions. More specifically, we shall look for models
where the proton is stable in the presence of either LV interactions (\ref{eq:LV}) 
or BV interactions (\ref{eq:BV}). We shall find that, without ever demanding it, the LV and BV terms 
are in fact naturally separated in the sense that the $U(1)'$ symmetry may allow
(\ref{eq:LV}) or (\ref{eq:BV}), but {\em not both} at the same time. 
This result, which we shall refer to as ``LV-BV separation'', 
is very general and relies only on the following three simple assumptions:
\begin{enumerate}
\item The MSSM Yukawa couplings are allowed by the $U(1)'$ gauge symmetry.
\item The $\mu$-problem is solved as in the UMSSM, namely 
the $U(1)'$ gauge symmetry forbids the original $\mu$ term (\ref{eq:mu}) 
while allowing the effective $\mu$ term (\ref{eq:mueff}).
\item There are no new exotic $SU(2)$ representations\footnote{In general, 
our results also hold in the presence of a certain number of 
additional pairs of Higgs doublets -- see Section \ref{sec:LVBVseparation}.} 
beyond the field content of the MSSM.
\end{enumerate}
The proof of the LV-BV separation is very simple and will be presented in Section \ref{sec:LVBVseparation}.

At this point, giving up on $R$-parity may seem like a rather steep price to pay.
After all, $R$-parity ensures that the lightest supersymmetric particle 
is stable and may provide a dark matter candidate. However, it is 
an under-publicized fact that $R$-parity by itself is not sufficient to 
stabilize the proton \cite{Weinberg,Ellis:1982wr,Ellis:1983qm,Harnik:2004yp}. 
While $R$-parity does prevent the proton from decaying
through the renormalizable operators (\ref{eq:LV}) and (\ref{eq:BV}), 
it still allows for potentially dangerous dimension five operators such as
\be 
W_5 = \frac{1}{\Lambda} C^L_{ijkl} Q_i Q_j Q_k L_l 
    + \frac{1}{\Lambda} C^E_{ijkl} U^c_i U^c_j D^c_k E^c_l
    + \frac{1}{\Lambda} C^N_{ijkl} U^c_i D^c_j D^c_k N^c_l\ ,
\label{eq:GUT}
\ee
which violate both lepton number\footnote{The lepton number of $N^c$ is given by $-1$
in the presence of an $H_2 L N^c$ term in the superpotential, 
which will be one of our assumptions later on (Section~\ref{sec:formalism}). 
Strictly speaking, $W_{\rm LV}$ of eq.~(\ref{eq:LV}) should also contain 
right-handed neutrino terms such as $N^cN^c$, $N^cN^cN^c$, and $SN^cN^c$ 
when a lepton number is assigned to $N^c$.} and baryon number.
Such operators are generically expected to appear at the cutoff scale $\Lambda$.
The problem with $R$-parity is that if, as expected, 
$\Lambda$ is on the order of the string scale or the Planck scale 
and the coefficients $C$ are of order one, 
the proton would still decay too fast \cite{Weinberg,Ellis:1982wr,Ellis:1983qm,Harnik:2004yp}.
In this sense, $R$-parity does not provide a complete and satisfactory 
solution to the proton decay problem\footnote{See, for instance Ref.~\cite{Sayre:2006en}, to see how grand unified theories can help with this problem.}.
The presence of the additional $U(1)'$ symmetry, however, offers new possibilities 
for dealing with the dangerous higher dimensional operators (\ref{eq:GUT}).
In fact we shall see that under the same three simple assumptions listed above, 
not only are the renormalizable LV and BV interactions (\ref{eq:LV}) and (\ref{eq:BV})
naturally separated, but also the dangerous non-renormalizable operators 
of the type (\ref{eq:GUT}) are automatically forbidden.
In this sense, in comparison to $R$-parity, the $U(1)'$ gauge symmetry 
may provide a more attractive alternative solution to the proton decay problem.

Our work is complementary to a number of studies in the literature 
which have already considered an extra non-anomalous $U(1)$ gauge symmetry 
in lieu of $R$-parity to address the proton stability 
problem \cite{Weinberg,Ellis:1982wr,Ellis:1983qm,Harnik:2004yp,U1proton,Font:1989ai,RPVnomu,CDM,Erler,AokiOshimo,Ma:2002tc}\footnote{For anomalous $U(1)$ approaches, see for example Ref.~\cite{Leontaris:1999wf} and references therein.}.
The more recent studies have adopted an even more economical 
approach, where the $U(1)'$ gauge symmetry is used to simultaneously solve 
both the $\mu$-problem and the proton stability problem \cite{CDM,Erler,AokiOshimo,Ma:2002tc}.
In those works the renormalizable $R$-parity violating interactions
(as well as the non-renormalizable interactions (\ref{eq:GUT}))
are forbidden by the $U(1)'$ symmetry\footnote{Previous 
studies \cite{RPVnomu} which considered 
$R$-parity violating interactions within the $U(1)'$ framework 
did not address the $\mu$-problem.}. 
The price to pay, however, was to allow for a relatively complicated 
spectrum, including e.g. $SU(2)_L$ exotics \cite{CDM,Erler},
several pairs of Higgs doublets ($N_H$) \cite{AokiOshimo}
or several singlet representations ($N_S$) \cite{AokiOshimo,Ma:2002tc}.
Even though our motivation here was to allow for either
LV or BV interactions, we have also analyzed cases 
where the $U(1)'$ symmetry forbids
all RPV operators of lowest dimensions.
Such examples are presented in appendices \ref{sec:nh4} and \ref{sec:nh3}. 
First in Appendix~\ref{sec:nh4} we consider the novel case of $N_H=4$,
while in Appendix~\ref{sec:nh3} we treat the case of $N_H=3$ which was
previously discussed in Ref.~\cite{AokiOshimo}. We shall show that in both of those
cases the nonlinear $U(1)'$ anomaly conditions
factorize and {\em all} anomaly conditions essentially reduce to linear 
constraints. Furthermore, the case of $N_H=3$, $N_S=3$ exhibits
an additional simplification: the quadratic and cubic $U(1)'$ anomaly conditions
are not independent, and we find a three-parameter 
class of anomaly-free solutions which generalize the single model 
found in Ref.~\cite{AokiOshimo}.

Previous studies found that the additional gauge symmetry usually also requires 
exotic fields for the cancellation of certain anomalies \cite{CDM,Erler,AokiOshimo,Ma:2002tc}. 
This tends to ruin the successful gauge coupling unification which is a hallmark 
of supersymmetry \cite{Wells}\footnote{Ref.~\cite{noexotics} considered
an UMSSM with family non-universal charges which was free of exotics.
However, in that case one can not write down Yukawa couplings for all 
fermions at tree level, and in Ref.~\cite{noexotics} non-holomorphic terms
were introduced in order to radiatively generate the problematic Yukawa couplings.}.
Here we do not require gauge coupling unification, and follow a 
bottom-up approach by introducing only the minimal set of
exotic fields (three vectorlike pairs of colored triplets $K_i$ and $K_i^c$, 
see Section~\ref{sec:A1}) 
required for anomaly cancellation. For simplicity,
we will also assume family universal $U(1)'$ charges for all MSSM fields,
including the right-handed neutrinos, 
but will let the exotics have family non-universal charges.

Our paper is organized as follows. In Section~\ref{sec:general}
we describe the general properties of our solutions. 
For this purpose, we shall only need to 
use the linear constraints on the $U(1)'$ charges following from the 
Yukawa-type couplings in the superpotential, plus the
$U(1)' - [SU(2)_L]^2$ anomaly condition from Section \ref{sec:A2}. 
We begin by introducing our formalism and notation in Section~\ref{sec:formalism}
and proceed to derive some of our main results in the remainder of 
Section~\ref{sec:general}. 
In Section~\ref{sec:LVBVseparation} we explicitly show the LV-BV separation, 
namely, that the renormalizable LV terms and BV terms can not coexist: 
if we allow for the LV terms (\ref{eq:LV}) in the superpotential, 
then the BV terms (\ref{eq:BV}) are automatically forbidden by the 
$U(1)'$ gauge symmetry, and vice versa. Then in Section~\ref{sec:HDO}
we extend our discussion to the case of the non-renormalizable RPV
terms such as (\ref{eq:GUT}) and show that those are absent as well.
In Section~\ref{sec:Maj} we derive a simple expression for the $U(1)'$ 
charge of the right-handed neutrino in terms of the $U(1)'$ charges of 
the other UMSSM fields, and discuss the origin of neutrino masses in our scenario.
Finally, in Section~\ref{sec:master} we present the general solution 
to the linear constraints discussed in Section~\ref{sec:formalism}
and then its specific form for the LV case or the BV case alone.
In Section~\ref{sec:anomalies} we discuss the remaining constraints 
on the $U(1)'$ charges arising from the absence of gauge anomalies.
We consider the anomaly conditions one at a time and discuss their implications 
for the model building to follow in the next three sections.
In Section \ref{sec:nh1} we present our simplest models 
($N_H=1$) with either LV or BV, but not both, type of interactions.
We summarize and conclude in Section~\ref{sec:conclusions}.
In Appendix~\ref{sec:nh4} (Appendix~\ref{sec:nh3}) we discuss 
models with $N_H=4$ ($N_H=3$) in which both types of
RPV terms are forbidden by the $U(1)'$ symmetry.
In Appendix \ref{app:nh1ns1}, we discuss a special case of 
$N_H = N_S = 1$ with an altered particle spectrum.

\section{General properties of the $U(1)'$ models}
\label{sec:general}

\subsection{Setup and Formalism}
\label{sec:formalism}

In the same spirit as the earlier works \cite{CDM,Erler,AokiOshimo,Ma:2002tc}, 
we consider the $U(1)'$-extended MSSM 
where both the $\mu$ term and the $R$-parity violating terms in the superpotential
are controlled by the $U(1)'$ gauge symmetry.
In contrast to previous studies along these lines, we shall not forbid
all renormalizable RPV terms from the very beginning. Instead, we shall 
in principle allow for the presence of either LV or BV terms in the superpotential.
We will not be particularly concerned whether the RPV terms 
(\ref{eq:LV}) and (\ref{eq:BV}) arise at the 
renormalizable level or through a higher dimensional operator. 
In fact, we shall find examples of both types below.
We shall then demonstrate that, as a result of the $U(1)'$ symmetry,
the proton is nevertheless still sufficiently stable,
even at the non-renormalizable level. Our result is quite general and 
relies only on our three simple assumptions listed in the Introduction. 

\begin{table}[tb]
\begin{center}
\begin{tabular}{||c||c|c|r|l||}
\hline\hline
Field & $SU(3)_C$ & $SU(2)_L$ & $U(1)_Y$         & $~U(1)'~$ \\
\hline\hline
$Q$     & $3$ & $2$ & $\frac{1}{6}$~~            & $~z[Q]$ \\
$U^c$   & $\overline 3$ & $1$ & $-\frac{2}{3}$~~ & $~z[U^c]$ \\
$D^c$   & $\overline 3$ & $1$ & $\frac{1}{3}$~~  & $~z[D^c]$ \\
$L$     & $1$ & $2$ & $-\frac{1}{2}$~~           & $~z[L]$ \\
$E^c$   & $1$ & $1$ & $1$~~                      & $~z[E^c]$ \\
$N^c$   & $1$ & $1$ & $0$~~                      & $~z[N^c]$ \\
$H_2$   & $1$ & $2$ & $\frac{1}{2}$~~            & $~z[H_2]$ \\
$H_1$   & $1$ & $2$ & $-\frac{1}{2}$~~           & $~z[H_1]$ \\
$S$     & $1$ & $1$ & $0$~~                      & $~z[S]$ \\
$K_i$   & $3$ & $1$ & $y[K_i]$                   & $~z[K_i]$ \\
$K^c_i$ & $\overline 3$ & $1$ & $-y[K_i]$        & $~z[K^c_i]$ \\
\hline\hline
\end{tabular}
\caption{Chiral fields in the model and their quantum numbers. 
$z[F]$ denotes the $U(1)'$ charge of a field $F$. 
In general, we consider $N_H$ pairs of Higgs doublets $H_1$ and $H_2$
with identical quantum numbers, and $N_S$ copies of SM Higgs singlets $S$.
\label{tab:SMcharges}}
\end{center}
\end{table}

To set up our discussion, in Table~\ref{tab:SMcharges} we list the particles of 
the UMSSM with their corresponding SM quantum numbers and 
$U(1)'$ charges. The first column lists the corresponding field, 
and the next two columns give its representation under $SU(3)_C$ and $SU(2)_L$.
The last two columns show the hypercharge $y[F]$ and the $U(1)'$ charge
$z[F]$ of a field $F$. In addition to the MSSM fields 
$Q$, $U^c$, $D^c$, $L$, $E^c$, $H_1$ and $H_2$,
we also include three right-handed neutrinos $N^c$. 
The Higgs singlet $S$ is introduced in order to generate
the effective $\mu$ term (\ref{eq:mueff}), and a successful solution 
to the $\mu$-problem requires that 
\be
z[S] = - z[H_1] - z[H_2] \ne 0\ .
\label{eq:zSnot0}
\ee
In what follows, we shall make repeated use of this equation which is nothing 
but the second of our three basic assumptions listed in the Introduction.
In general, we shall consider $N_H$ pairs of Higgs doublets $H_1$ and $H_2$
with identical quantum numbers, and $N_S$ SM Higgs singlets of type $S$.
The Abelian gauge symmetry $U(1)'$ is assumed to be broken at the TeV scale
where all Higgs fields ($S$, $H_1$ and $H_2$) get VEV's of that order. 
An effective $\mu$ term ($\mu_{\rm eff} = h \langle S\rangle$) is thus
dynamically generated at the TeV scale, completing the solution to the $\mu$-problem.
This is very similar to the case of the NMSSM, 
but having the $U(1)'$ gauge symmetry of the UMSSM has the additional advantage of
eliminating the domain wall problem associated with the discrete symmetry of the 
NMSSM \cite{domainwall}\footnote{In addition, quantum gravity effects may violate a 
global symmetry unless it is a remnant of a gauge symmetry \cite{KraussWilczek}.}.
As we mentioned 
earlier, a minimum set of vectorlike colored exotics $K_i$, $K_i^c$ $(i=1,2,3)$ 
is also required for anomaly cancellation (see Section~\ref{sec:A1}). At this point, 
the hypercharges of the exotics and the $U(1)'$ charges of all fields
listed in Table \ref{tab:SMcharges} are yet to be determined. 

In the remainder of this Section we shall analyze the main properties 
of our solutions, based on a limited set of linear constraints for the $U(1)'$ charges.
The remaining constraints will be analyzed in Section~\ref{sec:anomalies}.
We shall first list the set of relevant equations, and proceed to 
analyze them in the subsequent subsections.

In addition to (\ref{eq:mueff}), we also require that the $U(1)'$ symmetry
allows the usual Yukawa couplings in the superpotential
\be
W_{\rm Yukawa} = y^D_{jk} H_1 Q_j D^c_k + y^U_{jk} H_2 Q_j U^c_k 
               + y^E_{jk} H_1 L_j E^c_k 
               + y^N_{jk} \left( \frac{S}{\Lambda}  \right)^a H_2 L_j N^c_k \ .
\label{eq:Yukawas} 
\ee
Here capital letters denote the superfields of the MSSM whose
quantum numbers are listed in Table~\ref{tab:SMcharges}.
Because of the observed smallness of the neutrino masses, we have in general 
allowed neutrino Yukawa couplings to arise from a non-renormalizable operator 
suppressed by some high scale $\Lambda$ \cite{Cleaver:1997nj}. However, in principle 
we do not exclude the possibility of $a=0$. We discuss the 
possible appearance of a Majorana mass term for $N^c$ in Section \ref{sec:Maj}. 
The presence of the Yukawa terms (\ref{eq:Yukawas}) leads to the following
constraints 
\bea
~~~~Y_D&:&~~ z[H_1] + z[Q] + z[D^c]          = 0 \label{eq:YD} \\
~~~~Y_U&:&~~ z[H_2] + z[Q] + z[U^c]          = 0 \label{eq:YU} \\
~~~~Y_E&:&~~ z[H_1] + z[L] + z[E^c]          = 0 \label{eq:YE} \\
~~~~Y_N&:&~~ z[H_2] + z[L] + z[N^c] + a z[S] = 0 \label{eq:YN}\ . 
\eea
We supplement these with eq.~(\ref{eq:zSnot0}) which we write as
\bea
Y_S&:&~~ z[S] + z[H_1] + z[H_2]          = 0 \label{eq:YS} 
\eea
and the $U(1)' - [SU(2)_L]^2$ anomaly condition from Section \ref{sec:A2}
\be
A_2\,:\,~~ 9 z[Q] + 3 z[L] + N_H (z[H_1] + z[H_2]) + A_2({\rm exotics}) = 0\ . \label{eq:A2}
\ee
The set of 6 equations (\ref{eq:YD}-\ref{eq:A2}) is the starting point for
our analysis in the remainder of this Section. These 6 equations exactly
correspond to our three basic assumptions listed in the Introduction:
the existence of the Yukawa terms (\ref{eq:Yukawas}) is guaranteed by
eqs.~(\ref{eq:YD}-\ref{eq:YN}), the solution to the $\mu$-problem is implied by
eq.~(\ref{eq:YS}) and the absence of $SU(2)_L$ exotics among our particle 
content in Table~\ref{tab:SMcharges} simply means that there is no additional 
contribution to the $U(1)' - [SU(2)_L]^2$ anomaly and $A_2({\rm exotics})=0$
in eq.~(\ref{eq:A2}).

\subsection{LV-BV Separation}
\label{sec:LVBVseparation}

Starting with eqs.~(\ref{eq:YD}-\ref{eq:A2}) and
taking the linear combination $6Y_D+3Y_U-3Y_E+(N_H-3)Y_S-A_2$
gives the following constraint among the $U(1)'$ charges
\be
3 (z[U^c] + 2 z[D^c]) - 3 (2 z[L] + z[E^c]) + (N_H - 3) z[S] - A_2({\rm exotics}) = 0\ . \label{eq:guide}
\ee
We find this equation particularly useful both in illustrating one of our main points, 
as well as in categorizing the existing $U(1)'$ models in the literature.
Each term in eq.~(\ref{eq:guide}) corresponds to a particular physical situation:
\begin{enumerate}
\item The first term in eq.~(\ref{eq:guide}) represents the baryon number violating 
interactions of eq.~(\ref{eq:BV}). If this term is zero, BV interactions will be present 
in the model. In order to forbid (\ref{eq:BV}), one must have $z[U^c] + 2 z[D^c]\ne 0$,
which would require at least one of the remaining three terms in eq.~(\ref{eq:guide}) 
to be non-vanishing as well.
\item The second term in eq.~(\ref{eq:guide}) represents the lepton number violating 
interactions of eq.~(\ref{eq:LV}). If this term is zero, LV interactions will be present 
in the model. In order to forbid (\ref{eq:LV}), one must have $2 z[L] + z[E^c] \ne 0$,
which would require at least one of the remaining three terms in eq.~(\ref{eq:guide}) 
to be non-vanishing as well.
\item The third term in eq.~(\ref{eq:guide}) simply counts the number $N_H$ 
of Higgs doublet pairs in the model.
This term would vanish only if $N_H=3$, since the solution to the $\mu$-problem requires 
$z[S]\ne 0$ (see eq.~(\ref{eq:zSnot0})). 
\item The fourth term $A_2({\rm exotics})$ represents the contribution to the 
$U(1)' - [SU(2)_L]^2$ anomaly from states not listed in Table \ref{tab:SMcharges}.
It is a model-builder's choice whether this term is present or not. 
\end{enumerate}

Eq.~(\ref{eq:guide}) allows us to categorize the existing $U(1)'$ models according to
how many and which of these four terms are non-vanishing.
For example, Ref. \cite{AokiOshimo} forbids all renormalizable RPV terms, 
hence the first two terms in eq.~(\ref{eq:guide}) are both nonzero. In fact, they
cancel each other, since Ref.~\cite{AokiOshimo} assumes three pairs of Higgs doublets ($N_H=3$)
and no $SU(2)_L$ exotics, so that the last two terms in eq.~(\ref{eq:guide}) are zero. 
On the other hand, the models of Refs.~\cite{CDM,Erler} illustrate the case where
all four terms in eq.~(\ref{eq:guide}) are non-vanishing: those models 
also forbid RPV interactions, but contain $SU(2)_L$ exotics and have $N_H\ne 3$.
Finally, the models of Ref.~\cite{Ma:2002tc} have $N_H=1$ and no $SU(2)_L$ 
exotic representations, so they illustrate the intermediate case
where three terms in eq.~(\ref{eq:guide}) are non-vanishing.

According to our third basic assumption (see Introduction),
our approach will be to assume that there are no $SU(2)_L$ exotic representations 
so that $A_2({\rm exotics})=0$, in which case eq.~(\ref{eq:guide}) 
becomes
\be
3 (z[U^c] + 2 z[D^c]) - 3 (2 z[L] + z[E^c]) + (N_H - 3) z[S] = 0\ . \label{eq:guide3}
\ee
We shall be mostly interested in cases with $N_H \ne 3$, 
so that the third term in eq.~(\ref{eq:guide3}) is nonzero.
For simplicity, we shall concentrate on $N_H=1$ in Section~\ref{sec:nh1} 
(the case of $N_H=4$ is treated in Appendix~\ref{sec:nh4}).
Under those circumstances, eq.~(\ref{eq:guide3}) reveals that, at least at the renormalizable level,
the LV terms (\ref{eq:LV}) and the BV terms (\ref{eq:BV}) can not coexist 
(i.e.~the first two terms in eq.~(\ref{eq:guide3}) can not vanish simultaneously),
since we need at least one of them to cancel the non-vanishing third term 
proportional to $z[S]$. We refer to this mutual 
exclusion as the ``LV-BV separation''. The proton is then safe from decaying 
through renormalizable RPV interactions, even though $R$-parity is not present in the model.
Furthermore, one does not need {\em both} of the first two terms in eq.~(\ref{eq:guide3})
in order to cancel the third one -- only one of the first two terms will suffice.
Therefore we are free to consider models where either the first or the second 
term in eq.~(\ref{eq:guide3}) is zero and the corresponding RPV interactions are allowed.
For example, in the LV case, where $2 z[L] + z[E^c]=0$, eq.~(\ref{eq:guide3}) gives
\be
z[U^c] + 2 z[D^c] =  \left(1 - \frac{N_H}{3}\right) z[S] \ne 0
\label{eq:udd}
\ee
and the BV interactions (\ref{eq:BV}) are not allowed. Similarly, 
in the BV case, where $z[U^c] + 2z[D^c]=0$, eq.~(\ref{eq:guide3}) gives
\be
2 z[L] + z[E^c] = - \left(1 - \frac{N_H}{3}\right) z[S] \ne 0
\label{eq:lle}
\ee
and the LV terms (\ref{eq:LV}) are not allowed.
It is straightforward to see that the LV-BV separation also holds 
if the corresponding LV and BV terms arise at the 
non-renormalizable level -- in that case, there are extra contributions to
the right-hand side of eqs.~(\ref{eq:udd}) and (\ref{eq:lle})
which are integer multiples of $z[S]$, so that our argument 
still applies as long as $N_H=1$.

\subsection{Higher Dimensional Operators and Proton Decay}
\label{sec:HDO}

As we already mentioned in the Introduction, $R$-parity 
allows for potentially dangerous higher-dimensional operators 
like (\ref{eq:GUT}) which may still destabilize the proton.
The new $U(1)'$ gauge symmetry can now be used to
eliminate those as well \cite{CDM,Erler,AokiOshimo,Ma:2002tc}.
It is interesting to note that simply by making use of 
eqs.~(\ref{eq:YD}-\ref{eq:A2}), and
without specifying the further details of the model,
we can readily compute the $U(1)'$ charge of any such operator 
and test whether it is allowed or not.  
For example, the linear combination 
$Y_D+2 Y_U+Y_E+(\frac{N_H}{3}-2)Y_S-\frac{1}{3}A_2$ leads to
\be
2z[U^c]+z[D^c]+z[E^c]+\left(\frac{N_H}{3}-2\right)z[S]=0\ ,
\ee
which allows us to determine the $U(1)'$ charge of the $U^cU^cD^cE^c$
operator as
\be
U^cU^cD^cE^c~:~~~ 2z[U^c]+z[D^c]+z[E^c]=\left(2-\frac{N_H}{3}\right)z[S]\ .
\label{eq:zUUDE}
\ee
Since the solution to the $\mu$-problem already implies $z[S]\ne 0$
(see eq.~(\ref{eq:zSnot0})), this operator is forbidden, 
unless one allows for exactly 6 pairs of Higgs doublets in the model.
Similarly, the operator $QQQL$ is also absent, since 
its charge can be obtained from the linear combination
$\frac{1}{3}A_2-\frac{N_H}{3}Y_S$:
\be
QQQL~:~~~ 3z[Q]+z[L] = \frac{N_H}{3}z[S]\ .
\label{eq:zQQQL}
\ee
Because of eq.~(\ref{eq:zSnot0}), again it is clear that 
the $U(1)'$ symmetry does not allow this operator, since we 
already have at least one pair of Higgs doublets as in the MSSM.
Finally, one can obtain the $U(1)'$ charge of the operator
$U^cD^cD^cN^c$ from the combination
$2Y_D+Y_U+Y_N+(\frac{N_H}{3}-2)Y_S-\frac{1}{3}A_2$ as
\be
U^cD^cD^cN^c~:~~~ z[U^c]+2z[D^c]+z[N^c]=\left(2-a-\frac{N_H}{3}\right)z[S]\ .
\label{eq:zUDDN}
\ee
Since $a$ is an integer, we see that, in general, as long as the number $N_H$ of 
Higgs doublet pairs is not divisible by three, this operator is also forbidden. 
Even when $N_H$ is divisible by three, there will be only 
one special value of the integer $a$, namely $a=2-\frac{N_H}{3}$, 
which would allow the existence of this operator. Since $a$ must be positive, 
there are only two special cases that one should be worried about:
$(N_H=3,a=1)$ and $(N_H=6,a=0)$. The case $N_H=6$ is already disfavored by
(\ref{eq:zUUDE}), while in the case $N_H=3$ which we study in Appendix~\ref{sec:nh3}, 
we shall consider only the case $a=0$ as in Ref.~\cite{AokiOshimo}. 

To summarize, so far we have shown that in the simplest cases such as
$N_H=1,2,4,\cdots$ the conditions (\ref{eq:YD}-\ref{eq:A2}) 
are sufficient to rule out the dangerous dimension 5 operators 
(\ref{eq:GUT}) which simultaneously violate baryon and lepton number.
This is already an important advantage of our models compared to
the usual $R$-parity conserving scenario.
However, since in our approach we are allowing some of the dimension 4
LV or BV interactions, we should also check for potentially dangerous 
{\em pairs} of dimension 4 and dimension 5 operators, 
which may in general arise from either 
$F$-terms or $D$-terms. For the case of the MSSM, the problematic
combinations were identified in Ref.~\cite{IbanezRoss} as 
\be
\{L   Q   D^c, Q Q Q H_1\},
\{U^c D^c D^c, Q U^c E^c H_1\}, 
\{U^c D^c D^c, U^c D^{c\dagger} E^c\}, 
\{U^c D^c D^c, Q U^c L^\dagger\}\ .
\label{eq:pairs}
\ee
Using eqs.~(\ref{eq:YD}-\ref{eq:A2}), it is easy to 
derive the following relations between the $U(1)'$ charges of
the operators in each pair:
\bea
(z[L]+z[Q]+z[D^c])+(3z[Q]+z[H_1]) & = & \frac{N_H}{3}z[S]\ ,  \\
(z[U^c]+2z[D^c])+(z[Q]+z[U^c]+z[E^c]+z[H_1])    &=&\left(2-\frac{N_H}{3}\right)z[S]\ ,\\
(z[U^c]+2z[D^c])+(z[U^c]-z[D^{c}]+z[E^c])&=&\left(2-\frac{N_H}{3}\right)z[S]\ ,\\
(z[U^c]+2z[D^c])+(z[Q]+z[U^c]-z[L])   &=&\left(2-\frac{N_H}{3}\right)z[S]\ .
\eea
We see that all of the dangerous pairs of operators are 
forbidden by the $U(1)'$ symmetry, due to the condition (\ref{eq:zSnot0}).
(The case $N_H=6$ would in principle allow the last three pairs, but
$N_H=6$ was already disfavored by eq.~(\ref{eq:zUUDE}) and we shall 
not be considering it any further.)

So far we have shown that the proton is not destabilized by
the potentially dangerous pairs of operators constructed out 
of MSSM fields only. Since our models have additional 
fields present ($N^c$, $S$, $K_i$ and $K_i^c$) beyond those of the MSSM,
we still need to check that those extra fields do not 
give rise to dangerous pairs of operators analogous to (\ref{eq:pairs}).
We systematically checked all relevant combinations of dimension 4 
and/or dimension 5 operators 
involving $N^c$ and $S$ in addition to the usual MSSM fields,
and verified that all combinations which violate lepton number and baryon number are forbidden 
by the $U(1)'$ symmetry when $z[S] \ne 0$, 
and $\frac{N_H}{3}$ is not an integer\footnote{This statement is 
strictly true in the LV case. In the BV case the only potentially
troublesome pair of operators is $U^cD^cD^c$ and $N^cN^cN^cS$.
The latter has $U(1)'$ charge $(7-3a-N_H)z[S]$ (see eq.~(\ref{eq:BVsol}))
and is in principle allowed for the following three choices:
$\{a,N_H\}=\{0,7\},\{1,4\},\{2,1\}$. However, neither of these 
three options is a viable one: $N_H=7$ is incompatible with
the $A_3$ anomaly (see eq.~(\ref{eq:A3sol}) below);
$a=1$, $N_H=4$ is inconsistent with the $A_4$ anomaly
(see Appendix~\ref{sec:nh4}); while $a=2$
would imply too small neutrino masses.}.

It remains to discuss the effect of the colored exotics $K$, $K^c$ 
on proton decay. Since they are heavy, they can not 
appear among the proton decay products. However, they
may still mediate proton decay. It is more difficult to see 
that the proton is safe from such processes because the 
$U(1)'$ charges and hypercharges of the colored exotics 
are not determined by eqs.~(\ref{eq:YD}-\ref{eq:A2}).
One possible approach would be to choose the exotic 
hypercharges so that the lowest dimension operators 
coupling exotic quarks to the MSSM fields are absent 
\cite{Font:1989ai,AokiOshimo}. Here we shall consider 
a more general setup, where the hypercharges of the colored 
exotics in principle may allow couplings to the MSSM fields
(see Section~\ref{sec:A3}). The proof of proton stability
in that case will be presented in a separate publication \cite{Lee:2007qx} 
where we will discuss the discrete gauge symmetries \cite{IbanezRoss,Luhn1} 
encoded in our models.

\subsection{Majorana Neutrino Masses}
\label{sec:Maj}

Recent experiments show that the active neutrinos have masses.
There are different possibilities regarding the origin of neutrino masses:
e.g. Dirac neutrino masses may arise from the SM Higgs mechanism, 
and their smallness can be naturally explained through
a seesaw mechanism with heavy right-handed Majorana neutrinos \cite{seesaw}.
Other possibilities invoke extra dimensions \cite{XDneutrinomass} or
higher dimensional operators \cite{nuMassHDO}. Since we allow for
a neutrino Yukawa coupling (see eq.~(\ref{eq:Yukawas})), 
our models can readily accommodate Dirac type neutrinos.
In this subsection we investigate whether in addition to the
neutrino Yukawa coupling, one could write down a Majorana term
for the right-handed neutrinos, so that we can have some kind of a 
seesaw mechanism as well. 

Taking the linear combination $Y_E+Y_N-Y_S$ allows us to express the 
$U(1)'$ charge of the right-handed neutrinos $N^c$ as
\be
z[N^c]=-(2z[L]+z[E^c]) +(1-a)z[S]
= \left\{
\begin{array}{l}
(1-a) z[S] ~~~~~~~~~~\quad (\text{LV case})\ ;\\
(2-a-\frac{N_H}{3}) z[S] ~~\quad (\text{BV case})\ .\label{eq:zN} 
\end{array}
\right.
\ee
We see that in the BV case, lepton number violating terms involving 
the $N^c$ field (e.g. $N^cN^c$, $N^cN^cN^c$, and $SN^cN^c$) can not 
be generated, unless $\frac{N_H}{3}$ is an integer. Therefore when 
$N_H\ne 3,6,\cdots$, the LV-BV separation holds even in the presence of 
$N^c$ fields with lepton number $-1$.
While the BV case can then have only Dirac neutrino 
mass terms, the LV case may in general allow a Majorana neutrino mass 
term $N^cN^c$ whenever $a=1$. 
However, in the LV case the $SN^cN^c$ term has a $U(1)'$
charge of $(3-2a) z[S]$ and is not allowed. The active neutrinos 
of the LV case may also get their masses without the RH neutrinos 
through $f$-$\widetilde f$ loops involving the $\lambda$ and $\lambda'$ couplings, 
or through $\nu$-$\widetilde H_2^0$ mixing due to the 
$\mu'_{\rm eff} L H_2$ term in eq.~(\ref{eq:LV}) \cite{RPVMSSM}.

\subsection{General Solution to the Yukawa Constraints and the $A_2$ Anomaly}
\label{sec:master}

In this subsection we present the general solution to the constraints 
(\ref{eq:YD}-\ref{eq:A2}) and then specify its particular form 
separately for the LV case and the BV case. 

Since (\ref{eq:YD}-\ref{eq:A2}) are 6 constraints for 
9 variables, we find a three-parameter solution as
\be
\left(
\begin{array}{l}
z[Q]\\
z[U^c]\\
z[D^c]\\
z[L]\\
z[E^c]\\
z[N^c]\\
z[H_2]\\
z[H_1]\\
z[S]
\end{array}
\right)
\ =\
\frac{\ell}{3}\
\left(
\begin{array}{r}
-1\\
 1\\
 1\\
 3\\
-3\\
-3\\
 0\\
 0\\
 0
\end{array}
\right)
\ +\
h_1\
\left(
\begin{array}{r}
 0\\
 1\\
-1\\
 0\\
-1\\
 1\\
-1\\
 1\\
 0
\end{array}
\right)
\ +\
\frac{s}{9}\
\left(
\begin{array}{r}
  N_H\\
9-N_H\\
 -N_H\\
    0\\
    0\\
9(1-a)\\
  -9\\
   0\\
   9
\end{array}
\right)\ ,
\label{eq:sol}
\ee
where $\ell$, $h_1$ and $s$ are arbitrary coefficients.
The notation for those is suggestive of their interpretation:
$\ell=z[L]$, $h_1=z[H_1]$, and $s=z[S]$.

In the LV case, we have an additional constraint, e.g.
$2z[L]+z[E^c]=0$, which implies the relation $h_1=\ell$
and the solution (\ref{eq:sol}) becomes
\be
\left(
\begin{array}{l}
z[Q]\\
z[U^c]\\
z[D^c]\\
z[L]\\
z[E^c]\\
z[N^c]\\
z[H_2]\\
z[H_1]\\
z[S]
\end{array}
\right)
\ =\
2\ell\
\left(
\begin{array}{r}
-\frac{1}{6}\\
 \frac{2}{3}\\
-\frac{1}{3}\\
 \frac{1}{2}\\
-1\\
 0\\
-\frac{1}{2}\\
 \frac{1}{2}\\
 0
\end{array}
\right)
\ +\
\frac{s}{9}\
\left(
\begin{array}{r}
  N_H\\
9-N_H\\
 -N_H\\
    0\\
    0\\
9(1-a)\\
  -9\\
   0\\
   9
\end{array}
\right)\ .
\label{eq:LVsol}
\ee
Not surprisingly, we recognize in the first 
column vector on the right-hand side the hypercharge assignments 
of the UMSSM fields from Table~\ref{tab:SMcharges}.
Indeed, the constraints (\ref{eq:YD}-\ref{eq:YS})
arise from gauge-invariant operators, so clearly they will be
satisfied by the hypercharges of the UMSSM fields. What is more 
important at this point is the additional remaining degree of freedom 
represented by the second term in the right-hand side of 
eq.~(\ref{eq:LVsol}), which will allow us to find nontrivial 
solutions for the $U(1)'$ charges, different from the usual hypercharge.

In the BV case, the corresponding additional constraint 
$2z[U^c]+z[D^c]=0$ implies $h_1=\ell+(1-\frac{N_H}{3})s$
and the solution (\ref{eq:sol}) can be written as
\bea
\left(
\begin{array}{l}
z[Q]\\
z[U^c]\\
z[D^c]\\
z[L]\\
z[E^c]\\
z[N^c]\\
z[H_2]\\
z[H_1]\\
z[S]
\end{array}
\right)
& = &
\left( 2\ell - \frac{2}{3} N_H s\right) \
\left(
\begin{array}{r}
-\frac{1}{6}\\
 \frac{2}{3}\\
-\frac{1}{3}\\
 \frac{1}{2}\\
-1\\
 0\\
-\frac{1}{2}\\
 \frac{1}{2}\\
 0
\end{array}
\right)
\ +\
\frac{s}{3}\
\left(
\begin{array}{c}
       0\\
       6\\
      -3\\
     N_H\\
  -3-N_H\\
6-N_H-3a\\
      -6\\
       3\\
       3
\end{array}
\right)\ .
\label{eq:BVsol}
\eea
Just as in the LV case (\ref{eq:LVsol}), the usual hypercharges
appear as a particular solution to the constraints (\ref{eq:YD}-\ref{eq:A2}), 
but there is an additional class of solutions with nonzero $z[S]$,
so that in general our solutions will be a linear combination of these two classes.

\section{Anomalies}
\label{sec:anomalies}

Table \ref{tab:anomaly} summarizes the anomaly cancellation conditions for the 
$U(1)'$ charges of the fields in our model.
In this section, we investigate these anomaly cancellation conditions
one by one and discuss their implications for model building.

\begin{table}[bt]
\begin{center}
\begin{tabular}{|c|c|l|l|l|c|}
\hline
Identifier & Anomaly & ~~~~~~~~~~~~~~~~~~~~~~~~~~~~~ Equation \\
\hline
\hline
$A_1$ & $U(1)'$-$[SU(3)_C]^2$ & ${\rm tr} [z t^a t^b] = \frac{1}{4} \delta^{ab} \sum_q z = 0$ ~(color triplet fermions only) \\
$A_2$ & $U(1)'$-$[SU(2)_L]^2$ & ${\rm tr} [z \tau^a \tau^b] = \frac{1}{2} \delta^{ab} \sum_{f_L} z = 0$ ~(doublet fermions only) \\
$A_3$ & $U(1)'$-$[U(1)_Y]^2$ & ${\rm tr} [z y^2] = \sum_f z y^2 = 0$ \\
$A_4$ & $U(1)_Y$-$[U(1)']^2$ & ${\rm tr} [y z^2] = \sum_f y z^2 = 0$ \\
$A_5$ & $[U(1)']^3$ & ${\rm tr} [z^3] = \sum_f z^3 = 0$ \\
$A_6$ & $U(1)'$-$[{\rm gravity}]^2$ & ${\rm tr} [z] = \sum_f z = 0$ \\
\hline
\end{tabular}
\caption{Anomaly cancellation conditions for the $U(1)'$ charges of the particles 
in our model listed in Table~\ref{tab:SMcharges}. The first column lists a shorthand
identifier for each condition, which will be used throughout the text.
\label{tab:anomaly}}
\end{center}
\end{table}

\subsection{Anomaly $A_1$ ($U(1)' - [SU(3)_C]^2$)}
\label{sec:A1}

We begin with the mixed $U(1)' - [SU(3)_C]^2$ anomaly which we denote with $A_1$.
First we rederive the well known result that the presence of the Yukawa couplings
in the superpotential (\ref{eq:YD}-\ref{eq:YS}) requires 
exotic representations beyond those of the MSSM. Denoting the contribution
of such exotics to the $U(1)' - [SU(3)_C]^2$ anomaly by $A_1({\rm exotics})$
we can write $A_1$ as
\be
A_1~:~~~3 \left( 2 z[Q] + z[U^c] + z[D^c] \right) + A_1({\rm exotics}) = 0\ . \label{eq:A1def}
\ee
The first term is the contribution of the 3 generations of quarks in the MSSM, 
while the second term is the potential colored exotics contribution.
Now taking the linear combination $A_1-3Y_U-3Y_D+3Y_S$, we get
\be
A_1({\rm exotics}) = -3 z[S] \ ,
\label{eq:A1iszS}
\ee
which, in light of eq.~(\ref{eq:zSnot0}), shows the need for colored exotic 
representations \cite{CDM,Erler,AokiOshimo,Ma:2002tc}.

In this paper, we shall assume that the exotics are $N_K$ vectorlike 
pairs of chiral fields $K_i$ and $K^c_i$ so that they do not alter the anomaly cancellation 
conditions among the SM gauge groups. More specifically, we assume that
they are triplets and anti-triplets of $SU(3)_C$  with equal and opposite 
$U(1)_Y$ hypercharges $\pm y[K_i]$ (see Table~\ref{tab:SMcharges}). 
Perhaps most importantly, as already mentioned earlier in the Introduction,
we are assuming that the exotics which are needed to cancel the
$A_1$ anomaly are $SU(2)_L$ singlets, so that $A_2({\rm exotics})=0$.
With those assumptions, eq.~(\ref{eq:A1def}) becomes
\be
A_1'~:~~~ 3 \left( 2 z[Q] + z[U^c] + z[D^c] \right) 
+ \sum_{i=1}^{N_K} \left( z[K_i] + z[K^c_i] \right) = 0\ . \label{eq:A1exotic}
\ee

In order to avoid conflict with experiment, the exotic quarks $K_i$ and $K_i^c$
must be sufficiently heavy \cite{coloredExotic}. 
If their masses arise from an ordinary mass term
$K K^c$ in the superpotential, then $U(1)'$ invariance implies 
$A_1({\rm exotics}) = \sum_{i=1}^{N_K} \left( z[K_i] + z[K^c_i] \right) = 0$ 
and the $\mu$-problem can not be solved because of the conflicting requirements 
of eqs.~(\ref{eq:zSnot0}) and (\ref{eq:A1iszS}).
We therefore choose to generate masses for {\em all} colored exotics at the TeV scale, 
through superpotential couplings to the $S$ field:
\be
W_{\rm exotics}=h''_{ij} S K_i K^c_j\ .
\label{eq:exmass}
\ee
Assuming that the couplings in eq.~(\ref{eq:exmass}) are diagonal, we get
the following constraint among the $U(1)'$ charges of the exotics
\be
Y_{K_i}~:~~~z[S] + z[K_i] + z[K^c_i] = 0\ . 
\label{eq:YK}
\ee
Since $z[S]\ne 0$, this equation reveals that $K$ and $K^c$ do {\em not} carry equal and opposite 
$U(1)'$ charges, even though their hypercharges are equal and opposite 
($y[K_i] + y[K^c_i] = 0$).

Now taking the linear combination $A_1'-3Y_U-3Y_D+3Y_S-\sum_{i=1}^{N_K}Y_{K_i}$ gives
\be
(3-N_K)z[S]=0\ .
\ee
Combined with eq.~(\ref{eq:zSnot0}), this determines the number of exotic families as
\be
N_K = 3\ .
\ee
Notice that the $A_1$ anomaly did not impose any constraints 
on the $U(1)'$ charges themselves, but simply fixed the number of 
allowed representations in the model. We shall see that the same
phenomenon will take place when we consider some of the other anomaly 
conditions below. In the end, this will leave us with sufficient freedom 
to find sets of $U(1)'$ charges which will satisfy all of our model 
requirements.

\subsection{Anomaly $A_2$ ($U(1)' - [SU(2)_L]^2$)}
\label{sec:A2}

This anomaly condition was already introduced as eq.~(\ref{eq:A2}) in Section \ref{sec:formalism}.
With our assumption that all exotics in the model are $SU(2)_L$ singlets, it becomes
\be
A_2~:~~~ 9 z[Q] + 3 z[L] + N_H \left( z[H_1] + z[H_2] \right) = 0\ .
\ee

\subsection{Anomaly $A_3$ ($U(1)' - [U(1)_Y]^2$)}
\label{sec:A3}

In general, the $A_3$ anomaly condition is given by
\bea
9 (2 z[Q] y[Q]^2 + z[U^c] y[U^c]^2 + z[D^c] y[D^c]^2) + 3 ( 2 z[L] y[L]^2 + z[E^c] y[E^c]^2 ) && \nn \\
+ 3 \sum_{i=1}^{N_K} ( z[K_i] y[K_i]^2 + z[K_i^c] y[K_i^c]^2) + N_H (2 z[H_1] y[H_1]^2 + 2 z[H_2] y[H_2]^2 )&=& 0
\eea
where $y[F]$ is the $U(1)_Y$ hypercharge of a field $F$ as given in Table \ref{tab:SMcharges},
and we have omitted terms involving fields with vanishing hypercharge ($N^c$ and $S$).
Substituting the known hypercharges from Table~\ref{tab:SMcharges}
and using (\ref{eq:YK}), we can rewrite it as
\be
A_3:~~z[Q]+8z[U^c]+2z[D^c]+3z[L]+6z[E^c]+N_H\left(z[H_1] + z[H_2]\right)
-6 z[S] \sum_{i=1}^{N_K} y[K_i]^2 =0\,.
\ee
Now taking the linear combination $A_3+A_2-8Y_U-2Y_D-6Y_E+(8-2N_H)Y_S$
leads to the following simple constraint
\be
\left( 4-N_H-3 \sum_{i=1}^{N_K} y[K_i]^2 \right) z[S]=0\ .
\ee 
Because of condition (\ref{eq:zSnot0}), this uniquely reduces to 
\be
\sum_{i=1}^{N_K} y[K_i]^2 = \frac{1}{3} \left(4 - N_H\right)\ ,
\label{eq:A3sol}
\ee
where the hypercharges are normalized as in Table~\ref{tab:SMcharges}.
We see that, just as was the case for $A_1$, the anomaly cancellation condition $A_3$ 
did not provide an additional constraint on the $U(1)'$ charges, but instead only limits 
the number of Higgs doublet pairs $N_H$ and the choice for exotic hypercharges $y[K_i]$. 
Since the left-hand side of eq.~(\ref{eq:A3sol}) must be positive-definite 
and $N_H$ is an integer, there are
only four possible choices for the number of Higgs doublet pairs: $N_H=1,2, 3$ or 4,
that we need to consider.
The case of $N_H=3$ was already considered in Ref.~\cite{AokiOshimo}
and we shall revisit it again in Appendix~\ref{sec:nh3}. 
We shall also consider the case of $N_H=4$ in Appendix~\ref{sec:nh4}.
Our main interest, however, 
will be in the minimal case of $N_H=1$, which will be discussed below 
in Section \ref{sec:nh1}.

Having fixed the number of Higgs doublet pairs $N_H$, eq.~(\ref{eq:A3sol})
provides a guideline for choosing the hypercharges of the colored exotics.
Since the $A_1$ anomaly already required $N_K=3$ (see Section \ref{sec:A1}),
it is clear that a family universal choice with rational numbers is only possible for 
$N_H=3$, with $y[K_i] = \pm \frac{1}{3}$, or for 
$N_H=4$, with $y[K_i] = 0$. In the case of $N_H=1$ or $N_H=2$, 
one would have to choose exotic hypercharges in a family non-universal way.
In general, there are many possible choices, but here we shall limit ourselves to those where
the exotic hypercharges are the same (up to a sign) as the hypercharges of the
corresponding $SU(2)_L$ singlet, color triplet representations in the MSSM ($U^c$ and $D^c$):
\bea
N_H=1 &\Longrightarrow&  y[K_i]= \left\{ \pm\frac{1}{3}, \mp\frac{2}{3}, \mp\frac{2}{3} \right\}\ ,\label{eq:yinh1}  \\
N_H=2 &\Longrightarrow&  y[K_i]= \left\{ \pm\frac{1}{3}, \pm\frac{1}{3}, \mp\frac{2}{3} \right\}\ ,\label{eq:yinh2}  \\
N_H=3 &\Longrightarrow&  y[K_i]= \left\{ \pm\frac{1}{3}, \pm\frac{1}{3}, \pm\frac{1}{3} \right\}\ .\label{eq:yinh3} 
\eea
All three choices (\ref{eq:yinh1}-\ref{eq:yinh3}) satisfy the $A_3$ anomaly condition
(\ref{eq:A3sol}). The signs of the exotic hypercharges could be in general chosen arbitrarily.
We have limited ourselves to two cases -- with the upper signs in eqs.~(\ref{eq:yinh1}-\ref{eq:yinh3})
the exotics have the wrong quantum numbers to couple to the MSSM quarks and mediate proton decay.
In that case, however, the lightest exotic would be stable and may pose problems for cosmology.
This could be avoided, e.g. if the reheating temperature is very low, $T_{\rm RH} \lsim 100$ GeV, 
which may still be compatible with baryogenesis \cite{lowbaryogenesis}.
On the other hand, choosing the lower signs in eqs.~(\ref{eq:yinh1}-\ref{eq:yinh3})
allows the exotics to couple to the MSSM quarks, thus avoiding problems with cosmology.
Nevertheless, as we already discussed in Section~\ref{sec:HDO},
in that case the $U(1)'$ symmetry is sufficient to stabilize the proton.
We shall therefore allow for both sets of signs for the
exotic hypercharges in eqs.~(\ref{eq:yinh1}-\ref{eq:yinh3}).

\subsection{Anomaly $A_6$ ($U(1)' - [{\rm gravity}]^2$)}
\label{sec:A6}

The gravitational anomaly $U(1)' - [{\rm gravity}]^2$ is given as
\bea
A_6~:~~~9 (2 z[Q]+z[U^c]+z[D^c]) + 3 (2 z[L] + z[E^c] + z[N^c]) 
 && \nn \\
+ 2 N_H (z[H_1] + z[H_2]) + N_S z[S]  + 3 \sum_{i=1}^{N_K} (z[K_i] + z[K_i^c])  &=& 0\ , \label{eq:A6}
\eea
where $N_S$ is the number of Higgs singlets $S$ in the model.
Taking the linear combination
$A_6-9Y_U-9Y_D-3Y_E-3Y_N+(12-2N_H)Y_S-3\sum_{i=1}^{N_K}Y_{K_i}$,
we get
\be
\left( N_S-2N_H-3a+3\right) z[S]=0\ .
\ee
Because of eq.~(\ref{eq:zSnot0}), this implies
\be
N_S=2N_H+3a-3\ .
\label{eq:A6sol}
\ee
Once again, the anomaly condition did not constrain the 
$U(1)'$ charges, but just the number of representations.
The simplest possibility appears to be $N_H=1$, $a=1$, $N_S=2$,
and this is the case we shall investigate in Section~\ref{sec:nh1}.
Another example discussed in Appendix~\ref{sec:nh4} is $N_H=4$, $a=1$ and $N_S=8$.
Finally, $N_H=3$, $a=0$ and $N_S=3$ is the case considered in Ref.~\cite{AokiOshimo}
and below in Appendix~\ref{sec:nh3}.
We see that eq.~(\ref{eq:A6sol}) excludes the minimal 
(in the sense of total number $N_H+N_S$ of Higgs representations) 
possibility of $N_H = N_S = 1$ in our current setup\footnote{The 
gravitational anomaly $A_6$ was not taken into account in Ref.~\cite{Ma:2002tc},
which allowed building a model with $N_H = N_S = 1$.}.
However, this conclusion can be avoided with the addition of extra
SM singlet exotic fields. Appendix \ref{app:nh1ns1} 
provides a specific example of such a model with $N_H=N_S=1$.

\subsection{The Anomalies $A_4$ ($U(1)_Y - [U(1)']^2$) and $A_5$ ($[U(1)']^3$)}

The remaining anomaly conditions $A_4$ and $A_5$ are in general nonlinear equations for 
the $U(1)'$ charges:
\bea
A_4 &:& 9 ( 2 y[Q] z[Q]^2 + y[U^c] z[U^c]^2 + y[D^c] z[D^c]^2 ) + 3 (2 y[L] z[L]^2 + y[E^c] z[E^c]^2 ) \nn \\
&&+ 2N_H (y[H_1] z[H_1]^2 + y[H_2] z[H_2]^2 ) + 3 \sum_{i=1}^{N_K} (y[K_i] z[K_i]^2 + y[K_i^c] z[K_i^c]^2 )  = 0\ , 
\label{eq:A4def}\\
A_5 &:& 9 (2 z[Q]^3 + z[U^c]^3 + z[D^c]^3) + 3 (2 z[L]^3 + z[E^c]^3 + z[N^c]^3) \nn \\
&&+ 2N_H (z[H_1]^3 + z[H_2]^3) + N_S z[S]^3 + 3 \sum_{i=1}^{N_K} (z[K_i]^3+z[K_i^c]^3) = 0\ .
\label{eq:A5def}
\eea
Because of their nonlinearity, in the past $A_4$ and $A_5$ have typically been 
the stumbling blocks for finding anomaly-free solutions for the $U(1)'$ charges.
Here we shall show, however, that under our previous assumptions 
(\ref{eq:YD}-\ref{eq:A2}), {\em both} of these equations factorize --
each one is in fact proportional to $z[S]$ (which according 
to eq.~(\ref{eq:zSnot0}) is nonzero) so effectively we are able to reduce the power
of eq.~(\ref{eq:A4def}) and eq.~(\ref{eq:A5def}) by one\footnote{The factorization of
the $A_4$ and $A_5$ anomalies has been previously noticed in Ref.~\cite{Ma:2002tc}
for the specific case of $N_H=1$, $a=0$ and a particular set of exotics.}. For example, 
the $A_4$ anomaly reduces to a {\em linear} constraint among 
the $U(1)'$ charges. The easiest way to see this is to substitute the 
general solution (\ref{eq:sol}) into eq.~(\ref{eq:A4def}), which gives
\be
\frac{1}{3} s \left\{ (12 N_H - 36) h_1 + (7 N_H - 18) s - 12 \ell 
- 9 \sum_{i=1}^{N_K} y[K_i] (s + 2 z[K_i]) \right\} = 0\ .
\label{eq:A4fac}
\ee
Since $s \ne 0$, the expression within the curly brackets must vanish, which allows us to solve 
e.g. for one of the exotic charges $z[K_i]$ in terms of the other two as well as $s$, 
$h_1$ and $\ell$.

Similarly, substituting the general solution (\ref{eq:sol}) into eq.~(\ref{eq:A5def}), 
and using eq.~(\ref{eq:A6sol}), we get
\bea
s\, \Biggl\{
&-&3\biggl[ \bigl(3a+4N_H-12\bigr) h_1^2 -6a h_1\ell +(3a-4)\ell^2 \biggr]  \nn \\
&+& \biggl[ 3\bigl( 3a^2-6a-4N_H+12 \bigr) h_1 - \bigl( 9a^2-18a+2N_H\bigr)\ell   \biggr]\, s \nn \\
&-& \frac{1}{3}\biggl[ 9a^3-27a^2+18a-N_H^2+9N_H \biggr]\, s^2  
- 9 \sum_{i=1}^{N_K} z[K_i]  (s + z[K_i])
\Biggr\}=0\ .
\label{eq:A5fac}
\eea
Once again, since $s\ne0$, the expression within the curly brackets must vanish, 
which translates into only a {\em quadratic} constraint on the $U(1)'$ charges. 

As we shall see later in Appendix~\ref{sec:nh3}, a further 
drastic simplification of the above formulas (\ref{eq:A4fac}) and (\ref{eq:A5fac})
occurs for the case of $N_H=3$, $a=0$ and $y[K_i]=\pm \frac{1}{3}$, 
when the cubic anomaly completely factorizes, and 
effectively reduces to a linear constraint.
Furthermore, this linear constraint turns out to be 
equivalent to the constraint implied by eq.~(\ref{eq:A4fac}), so that
in effect the cubic anomaly condition is automatically satisfied
and in that case does not constrain the $U(1)'$ charges at all.

This completes our discussion of the anomaly cancellation conditions
involving the new $U(1)'$. To recapitulate, in Section \ref{sec:general}
we first considered the effect of the 6 constraints (\ref{eq:YD}-\ref{eq:A2}) 
on the $U(1)'$ charges of the 9 non-exotic fields in our model 
(see Table~\ref{tab:SMcharges}).
This resulted in the general three-parameter solution given by eq.~(\ref{eq:sol}).
Then in Section~\ref{sec:anomalies}, we studied the remaining\footnote{Recall 
that $A_2$ was already accounted for in Section \ref{sec:general}.}
5 anomaly cancellation conditions $A_1$, $A_3$, $A_4$, $A_5$ and $A_6$,
which involved 3 additional variables -- the $U(1)'$ charges $z[K_i]$
of the exotic fields $K_i$. We found that only 2 out of these 5 new 
conditions actually restrict the values of the $U(1)'$ charges, so that
there is still a lot of freedom remaining in the actual $U(1)'$ 
charge assignments. In the following  
we shall demonstrate this explicitly by presenting specific examples of
anomaly-free charge assignments which satisfy all of the model-building 
constraints considered so far. In Section~\ref{sec:nh1}
we shall find, as anticipated, that
there exist solutions which allow for either LV or BV, but not both.
Nevertheless, the proton will be stable in such models, as already 
discussed in Section~\ref{sec:HDO}, and the $\mu$-problem will be solved
by eq.~(\ref{eq:zSnot0}).

\section{Models with lepton or baryon number violation}
\label{sec:nh1}

In this section we shall concentrate on the simplest case of $N_H=1$.
In addition to the usual MSSM fields, the model also contains $N_S=2$ 
Higgs singlets $S_i$ and $N_K=3$ vectorlike pairs $(K_i,K^c_i)$ of exotic quarks
introduced to cancel the $A_1$ anomaly (see Section \ref{sec:A1}). 
The $R$-parity conserving part of the superpotential is given by
the combination of eqs.~(\ref{eq:mueff}), (\ref{eq:Yukawas}) and (\ref{eq:exmass}):
\bea
W_{\rm RPC}&=&y^D_{jk} H_1 Q_j D^c_k + y^U_{jk} H_2 Q_j U^c_k 
               + y^E_{jk} H_1 L_j E^c_k + y^N_{ijk} \frac{S_i}{\Lambda} H_2 L_j N^c_k \nonumber \\
           &+& h_i S_i H_2 H_1 + h''_{ijk} S_i K_j K^c_k\ .
\label{eq:UMSSMRPC} 
\eea
Recall that with $N_H=1$ and $N_S=2$, the $A_6$ anomaly condition
(\ref{eq:A6sol}) demands $a=1$, so that the neutrino Yukawa couplings arise
from a non-renormalizable operator as shown. 
We assume diagonal couplings of the exotics to $S$
(i.e. $z[K^c_i] = -z[K_i] - z[S]$) but off-diagonal terms may also 
exist if two or more exotic quarks have identical $U(1)'$ charges.
As discussed in the Introduction, the $\mu$-problem is solved through
an effective $\mu$ term $\mu_{\rm eff}=h_1\langle S_1 \rangle + h_2\langle S_2 \rangle$ by requiring
$z[S]\ne 0$. This forbids not only the original $\mu$ term (\ref{eq:mu}),
but also mass terms for the exotics ($K K^c$) and Higgs singlet self-couplings
$S$, $S^2$ and $S^3$.

The $R$-parity violating part of the renormalizable
superpotential of the UMSSM is
\be
W_{\rm RPV} = W_{\rm LV} + W_{\rm BV}\ ,
\label{eqn:UMSSMRPV}
\ee
where
\bea
W_{\rm LV} &=&\lambda_{ijk} L_i L_j E^c_k 
            + \lambda'_{ijk} L_i Q_j D^c_k 
            + h'_{ij} S_i H_2 L_j \ ,  \label{eq:UMSSMLV}  \\
W_{\rm BV} &=& \lambda''_{ijk} U^c_i D^c_j D^c_k \ .
\label{eq:UMSSMBV}
\eea
It is easy to see that the $U(1)'$
symmetry either simultaneously allows all three terms $\{ L L E^c, L Q D^c, S H_2 L \}$,
in which case $z[L]=z[H_1]$, or simultaneously forbids all three.  
In the LV case, therefore, we shall expect to have all three
terms appearing in eq.~(\ref{eq:UMSSMLV}) present.

A comment is in order regarding the possibility of a bare LV
$\mu' H_2 L_i$ term in the superpotential. Such a term is dangerous because
it will reintroduce a hierarchy problem ($\mu'$-problem) of the type we
originally intended to avoid. Indeed, the general solution (\ref{eq:sol})
in principle allows for this term. However, it is easy to see that 
in both the LV case and the BV case we are interested in, this term 
is absent and the $\mu'$-problem is solved in exactly the same way as 
the $\mu$-problem. For example, in the LV case the $U(1)'$ charge
of $H_2L_i$ from eq.~(\ref{eq:LVsol}) is $z[H_2L_i]=z[S]$ which is 
not vanishing because of condition (\ref{eq:zSnot0}). In the BV case,
from eq.~(\ref{eq:BVsol}) we get $z[H_2L_i]=(N_H-6)z[S]/3$. 
Since the case of $N_H=6$ was already discarded (see Section~\ref{sec:HDO}),
the $H_2L_i$ is again forbidden by the $U(1)'$ symmetry.
An effective $\mu'$ term will be nevertheless generated 
from the $S_iH_2L_j$ term in $W_{\rm LV}$, once the $U(1)'$ 
symmetry is broken by the VEV of $S$ at the TeV scale.

The $U(1)'$ symmetry is broken when $S$ gets a VEV $\langle S \rangle$ 
at the TeV scale. This generates the corresponding 
effective bilinear terms in the superpotential with coefficients
\be
\mu_{\rm eff} \equiv h_i \langle S_i \rangle\, , \qquad 
\mu'_{i, \rm eff} \equiv h'_{ji} \langle S_j \rangle\, , \qquad
m_{K, ij} \equiv h''_{kij} \langle S_k \rangle \ . 
\ee 
With the natural size of the couplings $\{h, h', h''\} \sim 1$,
the effective $\mu$ and $\mu'$ parameters as well as the masses 
of the exotic quarks $m_K$ are all of order a TeV.
With the effective bilinear terms, the superpotential of the UMSSM 
becomes similar to that of the MSSM.
First, the model predicts a new gauge boson, $Z'$, near the $U(1)'$ symmetry breaking scale:
\be
M_{Z'}^2 = g_{Z'}^2 \left( z[H_1]^2 v_1^2 + z[H_2]^2 v_2^2 + z[S]^2 v_{s1}^2 + z[S]^2 v_{s2}^2 \right)\ .
\ee
Here, $g_{Z'}$ is the $U(1)'$ gauge coupling constant, 
$v_i = \sqrt{2} \langle H_i \rangle$ (with $v_1^2 + v_2^2 \simeq 246^2 ~{\rm GeV}^2$), 
and $v_{si} = \sqrt{2} \langle S_i \rangle$.
The direct constraint on the mass of the $Z'$ comes from searches at the Tevatron 
in the dilepton channel ($Z' \to \ell^+ \ell^-$). The typical bound is 
$M_{Z'} > 600 \sim 900$ GeV, depending on the $U(1)'$ charges of the quarks 
and leptons \cite{Z2masslimit}. The VEV's of the Higgs doublets will also 
induce mixing between the $Z$ and $Z'$ gauge bosons.
If the $Z'$ is sufficiently heavy, this mixing is quite small, in accordance
with the experimental constraints from LEP (per mil level) \cite{LEPmixing}.
The supersymmetric partners of the $Z'$ and $S$ ($Z'$-ino and singlino) 
become extra components of the neutralinos.
The $S$ field gives one physical CP-even Higgs state, while the corresponding 
CP-odd Goldstone boson gets absorbed as the longitudinal component of the $Z'$ gauge boson.
For recent studies on phenomenology of the UMSSM, see Ref.~\cite{UMSSMphenomenology}.

We shall now present explicit
examples where the $U(1)'$ symmetry allows for either 
$W_{\rm LV}$ or $W_{\rm BV}$, but not both at the same time.
For simplicity, we assume the MSSM chiral fields ($Q, U^c, D^c, L, E^c, N^c$) 
to have family universal $U(1)'$ charges\footnote{Family 
non-universal $U(1)'$ charges in the SM quark sector may 
induce dangerous flavor changing neutral currents \cite{FCNCZ2}.
(On the other hand, such a flavor changing $Z'$ may provide an explanation 
of the discrepancies in rare $B$ decays \cite{ZprimeFCNC}.)}, but we 
allow family non-universal $U(1)'$ charges for the exotic quarks ($K_i, K_i^c$).
The hypercharges of the exotic quarks may be family non-universal as well.
In general, it is possible that there may be additional SM singlet fields 
which belong to the hidden sector, yet are charged under $U(1)'$ 
and thus contribute to the $A_5$ and $A_6$ anomalies. However,
our primary intention was simply to demonstrate that an anomaly-free $U(1)'$ 
can be used and is sufficient to achieve all of our goals 
outlined in the Introduction. Therefore, for concreteness and for simplicity, 
we shall assume only the field content listed in Table \ref{tab:SMcharges}.

In Table \ref{tab:U1charges}, we show several examples of anomaly-free charge 
assignments (up to an arbitrary normalization factor) 
for $N_H = 1$, $N_S=2$, $a=1$ and $y[K_i]=\{\frac{1}{3},-\frac{2}{3},-\frac{2}{3}\}$.
We have classified our examples in two groups: the first
five columns are LV models which allow for LV, but not BV terms in the superpotential,
while the remaining six columns are BV models 
which allow for BV, but not LV terms in the superpotential.
In LV models I-IV the LV terms appear already at the renormalizable 
level as in eq.~(\ref{eq:UMSSMLV}). 
In model V the terms of eq.~(\ref{eq:UMSSMLV})
appear at the non-renormalizable level
($SLLE^c$, $SLQD^c$ and $S^2H_2L$)
and in addition there are renormalizable LV terms
involving exotics, e.g. $NK_1K_1^c$ and $EK_2K_1^c$.
Similarly, BV models I-III already allow renormalizable BV couplings 
as in eq.~(\ref{eq:UMSSMBV}),
while BV models IV-VI allow only non-renormalizable
BV operators such as $QQD^{c\dagger}$, $QQQH_1$ and $H_1 H_2 U^c D^c D^c$.

\begin{table}[tbp]
\begin{center}
\begin{tabular}{||l||r|r|r|r|r||r|r|r|r|r|r||}
\hline\hline
          &  \multicolumn{5}{c||}{LV} & \multicolumn{6}{c||}{BV}   \\
\cline{2-12}
          &   I  &   II  &  III &   IV  &   V    &  I   &   II &   III   &   IV  &   V     & VI  \\
\hline
\hline
$z[Q]$    &  $1$ &   $3$ &  $3$ &   $3$ &  $4$ & $ 1$   &  $3$  &  $15$  &   $0$ &  $ 0$   &  $0$  \\
$z[U^c]$  &  $8$ &  $24$ & $24$ &  $24$ &  $5$ & $ 2$   &  $6$  &  $30$  &   $3$ &  $ 9$   &  $9$  \\
$z[D^c]$  & $-1$ &  $-3$ & $-3$ &  $-3$ & $-4$ & $-1$   & $-3$  &  $-15$ &   $0$ &  $ 0$   &  $0$  \\
$z[L]$    &  $0$ &   $0$ &  $0$ &   $0$ & $-9$ & $-2$   & $-6$  &  $-30$ &   $1$ &  $ 3$   &  $3$  \\
$z[E^c]$  &  $0$ &   $0$ &  $0$ &   $0$ &  $9$ & $ 2$   &  $6$  &  $30$  &  $-1$ &  $-3$   & $-3$  \\
$z[N^c]$  &  $0$ &   $0$ &  $0$ &   $0$ &  $9$ & $ 2$   &  $6$  &  $30$  &  $-1$ &  $-3$   & $-3$  \\
$z[H_2]$  & $-9$ & $-27$ &$-27$ & $-27$ & $-9$ & $-3$   & $-9$  &  $-45$ &  $-3$ &  $-9$   & $-9$  \\
$z[H_1]$  &  $0$ &   $0$ &  $0$ &   $0$ &  $0$ & $ 0$   &  $0$  &  $0$   &   $0$ &  $ 0$   &  $0$  \\
$z[S]$    &  $9$ &  $27$ & $27$ &  $27$ &  $9$ & $ 3$   &  $9$  &  $45$  &   $3$ &  $ 9$   &  $9$  \\
\hline
$z[K_1]$  & $-5$ & $-13$ &$-23$ & $-25$ & $-5$ & $-1$   & $-7$  &  $-17$ &  $-3$ &  $-7$   & $-5$  \\
$z[K_2]$  & $-2$ &  $-4$ & $-8$ &  $-7$ & $-5$ & $-1$   & $-4$  &  $-20$ &   $0$ &  $-1$   &  $1$  \\
$z[K_3]$  &  $1$ &   $2$ &  $1$ &  $-1$ & $-5$ & $-1$   & $-4$  &  $-11$ &   $0$ &  $ 2$   &  $1$  \\
$z[K^c_1]$& $-4$ & $-14$ & $-4$ &  $-2$ & $-4$ & $-2$   & $-2$  &  $-28$ &   $0$ &  $-2$   & $-4$  \\
$z[K^c_2]$& $-7$ & $-23$ &$-19$ & $-20$ & $-4$ & $-2$   & $-5$  &  $-25$ &  $-3$ &  $-8$   &$-10$  \\
$z[K^c_3]$&$-10$ & $-29$ &$-28$ & $-26$ & $-4$ & $-2$   & $-5$  &  $-34$ &  $-3$ & $-11$   &$-10$  \\
\hline\hline
\end{tabular}
\caption{Examples of anomaly-free $U(1)'$ charge assignments
for $N_H = 1$, $N_S=2$, $a=1$ and $y[K_i]=\{\frac{1}{3},-\frac{2}{3},-\frac{2}{3}\}$.
These $U(1)'$ charges can be scaled by an arbitrary normalization factor, 
as well as rotated by hypercharge (see text for details).
\label{tab:U1charges}}
\end{center}
\end{table}

A few comments are in order. First, each example in Table~\ref{tab:U1charges}
in fact corresponds to a whole family of solutions. This is because 
hypercharge itself also satisfies all of our requirements, including
the absence of mixed anomalies with $U(1)'$. Therefore,
each one of our solutions can be ``rotated'' by hypercharge in an arbitrary 
normalization. More specifically, if $z_0[F_i]$ is any particular 
solution from Table~\ref{tab:U1charges}, then a family of anomaly-free
$U(1)'$ charges is generated by the linear combination
\be
z[F_i] \ = \ \alpha\, z_0[F_i] + \beta \, y[F_i]\ ,
\label{eq:hyprot}
\ee
where $y[F_i]$ are the hypercharge assignments of our fields $F_i$ 
from Table~\ref{tab:SMcharges} and $\alpha$ and $\beta$ are arbitrary coefficients.
Therefore, the numerical values for the $U(1)'$ charges in our models are 
subject to fixing the convention for eq.~(\ref{eq:hyprot}). In Table~\ref{tab:U1charges}
we only listed examples which are {\em not} equivalent in the sense of eq.~(\ref{eq:hyprot}).

In spite of the freedom provided by eq.~(\ref{eq:hyprot}), the 
numerical values of the $U(1)'$ charges are important for phenomenology, 
as they determine the couplings of the particles in our model to the $Z'$.
For instance, our LV examples I-IV in Table~\ref{tab:U1charges}
are completely leptophobic, as they have $z[L]=z[E^c]=z[N^c]=0$. 
Under those circumstances, the standard collider bounds on the $Z'$ mass
are degraded, and a very light $Z'$ can be allowed.
However, this is not a general property 
of our LV models, since the hypercharge ``rotation'' (\ref{eq:hyprot})
could generate nonzero $U(1)'$ charges for $L$, $E^c$ and $H_1$.
On the other hand, $z[N^c]=0$ {\em is} a general property of LV models 
I-IV in this 
particular case ($a=1$), as already anticipated by eq.~(\ref{eq:zN}). 
Similarly, the vanishing entries for the $U(1)'$ charges of $Q$, $D^c$ and $H_1$
in our BV models, can also be rotated away from zero using eq.~(\ref{eq:hyprot}).

As we mentioned in Section~\ref{sec:A3}, we also consider the 
case where the exotic hypercharges have the opposite sign:
$y[K_i]=\{-\frac{1}{3},\frac{2}{3},\frac{2}{3}\}$.
The actual solutions for the $U(1)'$ charges that we find in that 
case are given simply by those of Table~\ref{tab:U1charges}, with the
replacement $z[K_i]\leftrightarrow z[K^c_i]$.
In general, this choice of $y[K_i]$ appears dangerous, 
since hypercharge alone would then
allow for LV and BV couplings involving exotic fields. However, we find that
due to the general phenomenon of LV-BV separation discussed in 
Section~\ref{sec:LVBVseparation}, the $U(1)'$ symmetry is still sufficient 
to prevent the simultaneous appearance of LV and BV couplings in the 
superpotential, and in all but one case (namely, BV-IV with 
opposite exotic hypercharge) the proton turns out to be stable 
\cite{Lee:2007qx}.

\section{Conclusions}
\label{sec:conclusions}

In this paper, we constructed a $U(1)'$-extended MSSM without $R$-parity,
where the extra non-anomalous $U(1)$ gauge symmetry plays the dual role 
of solving the $\mu$-problem and controlling the $R$-parity violating terms
(\ref{eq:LV}-\ref{eq:BV}). The $U(1)'$ gauge symmetry
provides a solid theoretical framework for discussing the phenomenology of
$R$-parity violation.
The most important implication 
of our models is the LV-BV separation: when the lepton 
number violating terms (\ref{eq:LV}) are allowed by the 
$U(1)'$ symmetry, the baryon number violating terms (\ref{eq:BV})
in the superpotential are automatically forbidden, and vice versa.
Within our approach, the dangerous dimension 5 operators 
such as $Q Q Q L$ or $U^c U^c D^c E^c$, which are allowed by 
$R$-parity and could still destabilize the proton, are also eliminated.
This presents a very minimal solution to the proton decay problem 
which is alternative to $R$-parity. We showed that the LV-BV separation
holds under very general circumstances. Perhaps the most stringent 
and least motivated was our assumption that there are no exotic
$SU(2)_L$ representations. While one can not judge the validity of this
assumption without knowledge of the fundamental theory at high energies,
it is certainly consistent with the principle of ``Occam's razor''.

While in our LV and BV examples the corresponding RPV couplings are allowed 
by the symmetries, the size of those couplings is still undetermined.
The experimental upper bounds on the individual RPV couplings range from
$10^{-3}$ for $\lambda$ to $10^{-7}$ for $\lambda''$. 
We do not consider such small values particularly fine-tuned, especially 
when compared to the Yukawa couplings of the first generation fermions in the SM.
In fact such small RPV couplings may naturally originate from
higher-dimensional operators, without modifying the analysis and the conclusions
of our paper \cite{Lee:2007qx}.

An interesting feature of our setup is that all LV terms 
($\lambda L L E^c$, $\lambda' L Q D^c$, $\mu'_{\rm eff} H_2 L$) 
must co-exist, as long as one of them is allowed. This is 
phenomenologically interesting since, for instance, the observation of a 
sneutrino resonance in an $s$-channel at hadron colliders 
such as the Tevatron and the LHC requires both $\lambda$ and $\lambda'$ couplings.
Besides the relation among the $R$-parity violating terms, 
our models also provide a connection between the phenomenology of 
$R$-parity violation and $U(1)'$ extensions of the MSSM.
In this sense, a potential discovery of a $Z'$ resonance
at the Tevatron or LHC would motivate searches for
$R$-parity violating SUSY signatures, and vice versa.

\begin{acknowledgments}

Our work is supported by the Department of Energy under Grants No. DE-FG02-97ER41029 and No. DE-FG02-97ER41022.
We are grateful to D. Demir, L. Everett, C. Luhn, and S. Nasri for helpful discussions.
\end{acknowledgments}

\newpage
\appendix

\section{Models with $N_H=4$}
\label{sec:nh4}

In this Appendix we briefly consider the case of $N_H=4$.
Again we shall choose $a=1$, which
fixes $N_S=8$ in accordance with eq.~(\ref{eq:A6sol}).
The exotic hypercharges are uniquely determined from
eq.~(\ref{eq:A3sol}) to be $y[K_i]=0$. For simplicity, in this Appendix 
we shall assume that the exotic quarks also have the same 
$U(1)'$ charges as well: $z[K_1]=z[K_2]=z[K_3]\equiv k$.
With those choices, the quadratic and cubic anomaly conditions
(\ref{eq:A4fac}) and (\ref{eq:A5fac}) can be rewritten as
\be
\frac{2}{3}\, s \left( 6 h_1 + 5 s - 6 \ell \right) = 0\ ,
\label{eq:A4facnh4}
\ee
\be
-\frac{1}{12}s\, \Biggl\{
\left( 6 h_1 + 5 s - 6 \ell \right) 
\left(42 h_1 + 7 s + 6 \ell \right) 
+ 9 \left(s+6k\right)\left(5s+6k\right) \Biggr\}=0\ .
\label{eq:A5facnh4}
\ee
Notice that taking into account eq.~(\ref{eq:A4facnh4})
eliminates the first term in the curly brackets in eq.~(\ref{eq:A5facnh4})
and the $A_5$ anomaly condition completely factorizes:
\be
-\frac{3}{4}\, s\, \left(s+6k\right) \left(5s+6k\right) =0\ .
\ee
This allows us to obtain explicitly a family of anomaly-free
solutions for the $U(1)'$ charges of the fields in Table~\ref{tab:SMcharges}.
It turns out that all of these solutions forbid {\em both}
the LV and BV terms, something which could not have been expected 
on the basis of eq.~(\ref{eq:guide3}) alone. 
Indeed, the $A_4$ anomaly constraint (\ref{eq:A4facnh4}) 
is inconsistent with the individual constraints for the
LV case ($h_1 = \ell$) and the BV case ($h_1 = \ell + (1 - \frac{N_H}{3}) s$)
which were derived earlier in Section \ref{sec:master}.
In either case, compatibility with eq.~(\ref{eq:A4facnh4}) demands $s = 0$, 
which is not allowed by the condition (\ref{eq:zSnot0}).

The factorized constraints (\ref{eq:A4facnh4}) and (\ref{eq:A5facnh4}) 
can now be solved rather easily and the general solution (\ref{eq:sol})
can be written as
\bea
\left(
\begin{array}{l}
z[Q]\\
z[U^c]\\
z[D^c]\\
z[L]\\
z[E^c]\\
z[N^c]\\
z[H_2]\\
z[H_1]\\
z[S]  \\
z[K]  \\
z[K^c]
\end{array}
\right)
& = &
-2\ell\
\left(
\begin{array}{r}
 \frac{1}{6}\\
-\frac{2}{3}\\
 \frac{1}{3}\\
-\frac{1}{2}\\
 1\\
 0\\
 \frac{1}{2}\\
-\frac{1}{2}\\
 0 \\
 0 \\
 0
\end{array}
\right)
\ +\
\frac{s}{18}\
\left(
\begin{array}{r}
  8\\
 -5\\
  7\\
  0\\
 15\\
-15\\
 -3\\
-15\\
 18\\
 18\rho\\
-18 (1+\rho)
\end{array}
\right)
\label{eq:solnh4}
\eea
where $\rho = -\frac{1}{6}$ ($\rho = -\frac{5}{6}$)
in the case of $s + 6 k = 0$ ($5 s + 6 k = 0$).
As expected, we obtain a two-parameter family of solutions -- one parameter 
($\ell$) corresponds to the usual hypercharge assignments while the 
second parameter ($s$) gives the nontrivial part of the $U(1)'$ solution.

\section{Models with $N_H=3$}
\label{sec:nh3}

In this Appendix we shall consider $U(1)'$ models with $N_H=3$,
$N_S=3$ and $a=0$, as in Ref.~\cite{AokiOshimo}. 
As we already saw in Section~\ref{sec:LVBVseparation},
in that case one should either simultaneously allow 
or simultaneously forbid the LV and BV terms (see eq.~(\ref{eq:guide3})). 
Furthermore, $N_H=3$ allows for family-universal hypercharges 
of the exotic quarks (see eqs.~(\ref{eq:A3sol}) and (\ref{eq:yinh3})).
We shall consider two possible values for the exotic hypercharges:
$y[K_i]=+\frac{1}{3}$ and $y[K_i]=-\frac{1}{3}$.
For simplicity, in this Appendix we shall again assume 
universal $U(1)'$ charges for the exotic quarks: $z[K_1]=z[K_2]=z[K_3]\equiv k$.
The $A_4$ anomaly (\ref{eq:A4fac}) can then be written as
\be
A_4~:~~~\left\{
\begin{array}{l}
-2s\left(3k+2\ell+ s\right)=0,~~~~{\rm for}~y[K_i]=\frac{1}{3}\ ;\\
 2s\left(3k-2\ell+2s\right)=0,~~~~{\rm for}~y[K_i]=-\frac{1}{3}\ ;
\end{array}
\right.
\label{eq:A4nh3}
\ee
while the $A_5$ anomaly condition is independent of $y[K_i]$
and reads
\be
A_5~:~~~-3s\left(3k+2\ell+ s\right)\left(3k-2\ell+2s\right)=0\ .
\label{eq:A5nh3}
\ee
We can see that $A_5$ completely factorizes into linear
polynomials which already appear in the expression for $A_4$.
Therefore, $A_5$ does not provide an additional restriction on the $U(1)'$ charges,
i.e.~$A_5$ will be automatically satisfied for {\em any} choice of $U(1)'$ charges
which is consistent with $A_4$. Since $A_4$ is already a linear relation, this
allows us to derive a three-parameter class of solutions which generalize
the single model found in Ref.~\cite{AokiOshimo}. For $y[K_i]=+\frac{1}{3}$, 
from eqs.~(\ref{eq:sol}) and (\ref{eq:A4nh3}) we find the general solution
\be
\left(
\begin{array}{l}
z[Q]  \\
z[U^c]\\
z[D^c]\\
z[L]  \\
z[E^c]\\
z[N^c]\\
z[H_2]\\
z[H_1]\\
z[S]  \\
z[K]\\
z[K^c]
\end{array}
\right)
\ =\
\ell\
\left(
\begin{array}{r}
-1\\
-1\\
 1\\
 1\\
-1\\
-3\\
 2\\
 0\\
-2\\
 0\\
 2
\end{array}
\right)
\ +\
h_1\
\left(
\begin{array}{r}
 0\\
 1\\
-1\\
 0\\
-1\\
 1\\
-1\\
 1\\
 0\\
 0\\
 0
\end{array}
\right)
\ +\
k\
\left(
\begin{array}{r}
 -1\\
 -2\\
  1\\
  0\\
  0\\
 -3\\
  3\\
  0\\
 -3\\
  1\\
  2
\end{array}
\right)\ ,
\label{eq:solnh3Ypos}
\ee
in terms of the $U(1)'$ charges $\ell\equiv z[L]$, $h_1\equiv z[H_1]$ and 
$k\equiv z[K_i]$. This solution is anomaly-free and satisfies all of the
constraints discussed in Sections \ref{sec:general} and \ref{sec:anomalies}.
As a special case, it also contains the solution found in Ref.~\cite{AokiOshimo},
which we recover by imposing $8\ell=-7h_1=-7k$. For example, $\ell=\frac{7}{12}$, 
$h_1=k=-\frac{2}{3}$, gives
\be
z\left[Q,U^c,D^c,L,E^c,N^c,H_2,H_1,S,K,K^c\right]=
\left\{
\frac{1}{12},
\frac{1}{12},
\frac{7}{12},
\frac{7}{12},
\frac{1}{12},
-\frac{5}{12},
-\frac{1}{6},
-\frac{2}{3},
\frac{5}{6},
-\frac{2}{3},
-\frac{1}{6}
\right\}
\label{eq:Aokimodel}
\ee
which is exactly the charge assignment in the model of Ref.~\cite{AokiOshimo}.
In addition to our requirements listed in Sections \ref{sec:general} and \ref{sec:anomalies},
Ref.~\cite{AokiOshimo} demanded the presence of a Majorana mass term $SN^cN^c$ 
in the superpotential. This would imply the constraint $8\ell-2h_1+9k=0$, 
which still leaves us with a two-parameter class of solutions
\be
\left(
\begin{array}{l}
z[Q]  \\
z[U^c]\\
z[D^c]\\
z[L]  \\
z[E^c]\\
z[N^c]\\
z[H_2]\\
z[H_1]\\
z[S]  \\
z[K]\\
z[K^c]
\end{array}
\right)
\ =\
\ell\
\left(
\begin{array}{r}
-1\\
 3\\
-3\\
 1\\
-5\\
 1\\
-2\\
 4\\
-2\\
 0\\
 2
\end{array}
\right)
\ +\
\frac{k}{2}\
\left(
\begin{array}{r}
 -2\\
  5\\
 -7\\
  0\\
 -9\\
  3\\
 -3\\
  9\\
 -6\\
  2\\
  4
\end{array}
\right)
\ee
as a generalization of eq.~(\ref{eq:Aokimodel}).

For completeness, we shall also consider the other possible sign of the 
exotic hypercharges: $y[K_i]=-\frac{1}{3}$, since in that case $A_5$ is also
automatically satisfied due to its factorization (\ref{eq:A5nh3}), 
which makes it easy to
obtain another class of solutions satisfying all Yukawa constraints and 
all anomaly cancellation conditions. Putting together eqs.~(\ref{eq:A4nh3}) and 
(\ref{eq:sol}), we find
\be
\left(
\begin{array}{l}
z[Q]  \\
z[U^c]\\
z[D^c]\\
z[L]  \\
z[E^c]\\
z[N^c]\\
z[H_2]\\
z[H_1]\\
z[S]  \\
z[K]\\
z[K^c]
\end{array}
\right)
\ =\
\ell\
\left(
\begin{array}{r}
 0\\
 1\\
 0\\
 1\\
-1\\
 0\\
-1\\
 0\\
 1\\
 0\\
-1
\end{array}
\right)
\ +\
h_1\
\left(
\begin{array}{r}
 0\\
 1\\
-1\\
 0\\
-1\\
 1\\
-1\\
 1\\
 0\\
 0\\
 0
\end{array}
\right)
\ +\
\frac{k}{2}\
\left(
\begin{array}{r}
 -1\\
 -2\\
  1\\
  0\\
  0\\
 -3\\
  3\\
  0\\
 -3\\
  2\\
  1
\end{array}
\right)\ .
\label{eq:solnh3Yneg}
\ee
Unfortunately, this class of models does {\em not} 
solve the proton decay problem: as can be seen from
eq.~(\ref{eq:solnh3Yneg}), the $U(1)'$ symmetry still allows
$R$-parity violating couplings involving exotic fields, e.g.
$U^cD^cK^c$ and $LQK^c$.

\section{Models with $N_H = N_S = 1$}
\label{app:nh1ns1}

We have already seen that the $A_6$ anomaly condition 
(\ref{eq:A6sol}) restricts the number of Higgs representations
$N_H$ and $N_S$. As we mentioned in Section \ref{sec:A6}, the
minimal case of $N_H = 1$, $N_S = 1$ is not allowed within 
the model we have discussed so far. However, the constraint 
(\ref{eq:A6sol}) varies with the particle spectrum,
and here we provide an example with a slightly altered spectrum
which can allow $N_H = 1$, $N_S = 1$.
We simply add another SM singlet field $X$ with superpotential
\be
W_{X} = \frac{\xi}{2} S X X\ ,
\ee
so that the $U(1)'$ charge of $X$ is given by $z[X] = -\frac{1}{2} z[S]$.
The general solution (\ref{eq:sol}) is then rewritten as
\be
\left(
\begin{array}{l}
z[Q]\\
z[U^c]\\
z[D^c]\\
z[L]\\
z[E^c]\\
z[N^c]\\
z[H_2]\\
z[H_1]\\
z[S]\\
z[X]
\end{array}
\right)
\ =\
\frac{\ell}{3}\
\left(
\begin{array}{r}
-1\\
 1\\
 1\\
 3\\
-3\\
-3\\
 0\\
 0\\
 0\\
 0
\end{array}
\right)
\ +\
h_1\
\left(
\begin{array}{r}
 0\\
 1\\
-1\\
 0\\
-1\\
 1\\
-1\\
 1\\
 0\\
0
\end{array}
\right)
\ +\
\frac{s}{9}\
\left(
\begin{array}{r}
  N_H\\
9-N_H\\
 -N_H\\
    0\\
    0\\
9(1-a)\\
  -9\\
   0\\
   9\\
-9/2
\end{array}
\right)
\ee
with no additional free parameters.

The new $X$ particles will modify the anomaly conditions 
$A_6$ ($U(1)'-[\rm{gravity}]^2$) and $A_5$ ($[U(1)']^3$) which 
get additional contributions of $N_X z[X]$ and $N_X z[X]^3$, respectively.
Then eq.~(\ref{eq:A6sol}) is modified as
\be
N_S = 2 N_H + 3 a - 3 + \frac{1}{2} N_X
\ee
where $N_X$ is the number of families of the $X$ fields.
$N_H = 1$, $N_S = 1$ is now allowed with $a = 0$, $N_X = 4$.
As an existence proof, we provide an example of an anomaly-free LV model of this 
category with $y[K_i]=\{\frac{1}{3}, -\frac{2}{3}, -\frac{2}{3}\}$:
\bea
&&z\left[Q,U^c,D^c,L,E^c,N^c,H_2,H_1,S,K_1,K_2,K_3,K^c_1,K^c_2,K^c_3,X\right] \nn \\
&&= \left\{
4, 8, 2, -6, 12, 18, -12, -6, 18, -6, -3, -15, -12, -15, -3, -9
\right\} .
\eea



\end{document}